# Permeation of Low-Z Atoms through Carbon Sheets: Density Functional Theory Study on Energy Barriers and Deformation Effects

Stefan E. Huber*, Andreas Mauracher and Michael Probst*

Institute of Ion Physics and Applied Physics, University of Innsbruck, Technikerstraße 25, 6020 Innsbruck, Austria
* Corresponding authors. Stefan E. Huber: E-mail: s.huber@uibk.ac.at, phone: +43 512 507 6247, Fax: +43 512 507 2922. Michael Probst: E-mail: Michael.probst@uibk.ac.at, phone: +43 512 507 6260, Fax: +43 512 507 2922.

**Abstract**
Energetic and geometric aspects of the permeation of low-Z atoms through graphene sheets are investigated. Energy barriers and deformations are calculated via density functional theory for the permeation of H, He, Li and Be atoms at several surface sites and at a hollow site for atoms B, C, O and Ne atoms. Graphene is modeled by large planar polycyclic aromatic hydrocarbons and the convergence of both energy barriers and deformation curves with increasing size of these hydrocarbons is investigated. Effective energy curves are summarized for the atoms under consideration in three different interaction regimes realized different geometrical constraints. In addition to the bare graphene model, the interaction between low-Z atoms and 100% hydrogenated coronene as a model for graphane is also investigated. The adiabatic barriers range from about 5 eV (1 eV = 1.602 x $10^{-19}$ J) for H to about 20 eV for Ne. Facilitation of the permeation by temporary chemical bonding is observed for O and C and for B and Be when interacting with hydrogenated coronene. The results are in good agreement with existing experimental and theoretical work and reflect the essential physics of the dynamics of H bombardment of graphite (0001) surfaces. Implications for modeling chemical sputtering of graphite in a mixed-material scenario as it will be the case in the next step fusion experiment ITER and potential building are discussed.

## 1. Introduction

Many critical issues in present thermo-nuclear fusion research are closely related to the topic of plasma wall interaction [1]. For example, in order to predict if the constraints on tritium retention are within the limits, the interaction details of H (and deuterium and tritium, denoted as D and T, respectively) atoms and their ions with the wall materials should be known. These issues are even more complicated in ITER (International Thermonuclear Experimental Reactor) where several materials are planned to be used, in one scenario graphite for the divertor plates, tungsten in the upper divertor and dome and Be in the main wall.
In such a tokamak the so-called scrape-off layer (SOL) plasma interacts with the plasma facing components (PFCs). The electronic temperature of the SOL located outside the last closed flux surface is much lower than that of the core plasma where the fusion process shall be maintained. The bombardment of the walls by H atoms and ions gives rise to the pollution of the plasma by impurities expelled from the PFCs that can penetrate far into the plasma through the surfaces of the magnetic field. The divertor region acts as power and particle exhaust in order to maintain a pure core plasma [2]. For this region in ITER, carbon fiber composite is envisaged, amongst other choices. In addition to ubiquitous plasma species as H (or D and T) and the fusion product He, PWI gives rise to impurities of C, Be and W [2-5]. Furthermore, techniques like wall conditioning [6-12] and impurity seeding [13-17] generate an even greater variety of plasma impurities consisting of e.g. Li, B, N, O, Si, Ne, Ar, Xe. Understanding the interaction of these impurities with graphite surfaces is therefore important if one wants to control the PWI processes in a mixed-material fusion device such as ITER.

Because of the abundance of H, D and T, mainly the interaction of these particles with carbon-based materials has been studied extensively by computational techniques like molecular dynamics (MD) simulations [18-23] and quantum-chemical methods [24-26]. Besides that, there exist various studies of other atom/molecule/ion-surface interactions possibly relevant for magnetic fusion [27-31].

Our work is of the quantum chemical type. We performed quantum mechanical DFT calculations of energy profiles and geometric features of atom/graphene systems. Such profiles explain certain characteristics of the systems, especially the different behavior depending on size and electronic structure of the particles. To our knowledge, there has never been an overall atomistic investigation of the interaction and transmission of atoms with and through graphene sheets.

Energy profiles can also be used as effective energy functions that are needed in classical MD simulations to describe the interactions between particles. These potentials are typically derived from equilibrium situations [32], but are often of poor accuracy when applied to non-equilibrium processes and, especially, systems under extreme conditions [28]. There is much need for robust reactive analytical potentials of any level of sophistication.

Graphite is a multilayer consisting of two-dimensional graphene sheets. Thus, the (mechanistic) interaction of atoms and/or molecules with graphite surfaces can to some extent be modeled by their interaction with graphene [33]. This is due to the fact that in ideal graphite the individual graphene sheets are well separated by 3.35 Å and are held together by a weak dispersion force. Especially the energetics and deformation upon mechanistic bombardment can be expected to be negligibly influenced by the three dimensional structure of the graphite crystal though this is for sure not the case for e.g. electronic properties [34], which, however, are outside the scope of this work. Physisorption and chemisorption of H and of a variety of other atoms on graphite surfaces or on graphene has already been studied by DFT methods, see e.g. [35-37]. In this work, we do not focus on these processes but on the permeation of various fusion-relevant atoms through graphene. We investigate the energy barriers arising from the repulsive forces between valence electrons as well as the deformation of the graphite surface in the vicinity of the permeation site. For the light atoms H, He, Li, and Be we investigate permeation at three surface sites: The 'hollow site' in the center of one aromatic ring; the 'bridge site', above the midpoint of a C-C bond; and the 'top site' on top of one C atom. For the higher-Z atoms B, C, O and Ne, we only investigate permeation at the hollow site.

In tokamaks, interaction of high-energy ions with carbon-based PFCs is at the origin of physical sputtering whereas low energy (H) atoms are responsible for the chemical erosion of carbon surfaces. The present work deals only with neutral atomic species. Furthermore the energy surfaces are by definition time-independent although they drive the dynamics, and therefore they are identical for all isotopes of one element.

After briefly summarizing the numerical methods in section 2, we discuss the convergence properties of our resulting energy barriers and PAH deformation depending on the size of the cluster used to model the graphite (0001) surface in section 3.1. We continue with a discussion of our results on the adiabatic permeation of low-Z atoms H to C, O and Ne through graphene models at the hollow site in section 3.2. This is followed by analogous analyses of bridge and top sites in sections 3.3 and 3.4 for the atoms H to Be. In sections 3.5 and 3.6 we discuss the different interaction regimes and the interaction of atoms with hydrogenated hydrocarbons as models for graphane [38]. Finally, in section 4, a summary is given.

## 2. Method

### 2.1 Cluster Model

We use a cluster approach where graphite is modeled by polycyclic aromatic hydrocarbons (PAHs). This avoids some problems associated with periodic calculations but it requires that the cluster model has the same or similar properties as the extended system. We discuss this issue in section 3.1. The PAHs can be thought as small pieces of graphene sheets with the free valences of the dangling bonds saturated by H. Vice versa, a graphene sheet can be interpreted as an infinite PAH molecule. Successful utilization of PAH molecules in modeling graphite surfaces has been reported earlier [33]. The PAHs used in our work are (i) benzene with six carbon atoms, (ii) anthracene with 14 C atoms, (iii) pyrene with 16 C atoms, (iv) triphenyl with 18 C atoms, (v) coronene with 24 C atoms, (vi) circumpyrene with 42 C atoms and (vii) circumcoronene with 54 C atoms (figure 1).

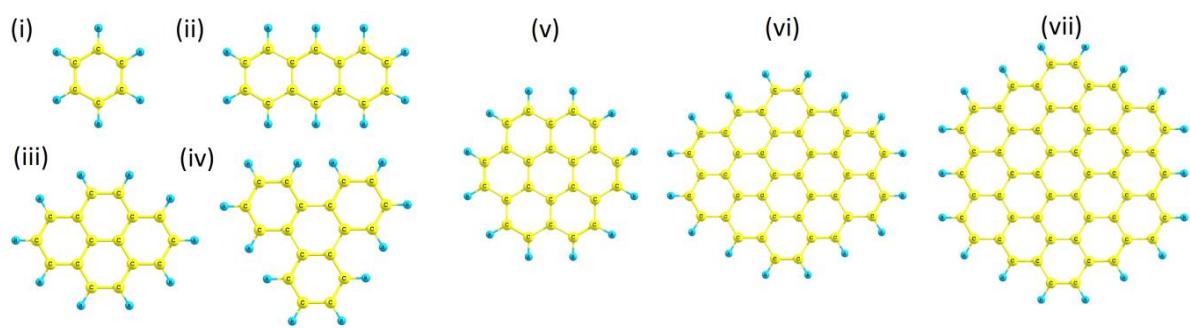

**Figure 1:** PAH molecules used as graphene models.

The interaction energy between these PAH molecules and the low-Z atoms is calculated by the density functional methods described in the next paragraph as a function of the distance z between the atom and the undisturbed PAH plane (z=0). This is done numerically by varying z in steps of 0.2 Å, where 1 Å = $10^{-10}$ m.

We considered three limiting cases: (a) the 'adiabatic' situation where at each distance of the approaching atom from the surface the PAH geometry is subject to full relaxation except for the H atoms at the PAH border, which are constrained to the original unperturbed molecular plane and define the value of z. This results in a bending of the cluster surface and a stretching or shortening of individual C-C bonds. As a measure of the amount of deformation we took the displacement of C atoms from the original PAH plane and the change of the C-C bond length in the vicinity of the permeation site. An illustration of the adiabatic case is depicted in figure 2. (b) differs from the 'adiabatic' situation in that the C atoms remain restricted to the original molecular plane. They can only move inside this plane in order to reduce the energy of the system. Finally, in (c) the PAH molecule remains rigid in its original equilibrium geometry.

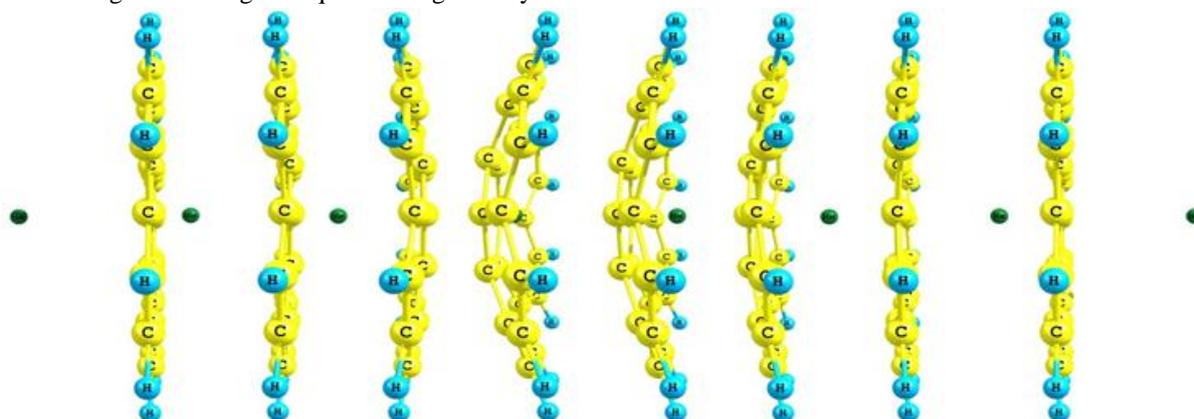

**Figure 2:** Illustration of the methodology used to calculate energy barriers and deformation effects in case of adiabatic permeation of He through coronene. The He atom moves from right to left.

Energies and geometries were obtained from DFT calculations with the B3LYP [39] and PBE0 [40] functionals. Both are hybrid functionals and include a mixture of Hartree-Fock exchange and DFT exchange-correlation. The widely used B3LYP functional has been constructed semi-empirically by fitting its three parameters to experimental data. The parameter-free PBE0 functional is based on the fulfillment of a number of physical constraints. The independent and complimentary foundation makes it interesting to use both functionals and to compare their results. A well-known deficiency of such density functionals is their inability to describe dispersion interactions. This would become important when one is, for example, interested in the shallow van-der-Waals attraction between surfaces and weakly interacting atomic species such as He and Ne. Even if the neglect of dispersion is not expected to be an issue for the energy barriers we are mostly interested in, we have recalculated 10 potential energy curves with the ωB97XD functional which accounts for dispersion interactions [41]. For simplicity, we do not include this in the following discussion but just mention that, comparing with the B3LYP and PBE0 results, we found similar shapes of the energy profiles and similar values for their energy maxima for all three. The deviations between the two types of functionals are found to be within 10%, noting that the ωB97XD interaction energies are slightly larger than the ones from B3LYP and PBE0. In the DFT calculations the electron densities are expanded in basis sets. Test calculations with large and small basis sets show that the split-valence basis sets 3-21G [42] and 6-31G [43] we used here overestimate the barrier maxima by up to 10% with respect to larger basis sets. Thus, we find that the calculated energy barriers are too low due to the neglect of dispersion and too high due to the use of the small basis sets. However, both effects give rise to deviations within an accuracy of 10%, such that one can expect that most of the error cancels. To be on the safe side anyway, we conclude that the accuracy of our method to estimate the energy barriers resulting from permeation is at least within 10% at the energy maximum.

All calculations have been performed with the Gaussian 09 software [44].

**2.2 Energy decomposition**

The planar hydrocarbon molecule becomes distorted when the impinging atom approaches. In most cases, it bends away from the atom, outwards of the original plane in order to reduce the repulsion between the electrons. The atom-PAH interaction energy ΔE is then

$\Delta E = E_{tot}(A\text{-}PAH) - E_{tot}(PAH) - E_{tot}(A)$

where $E_{tot}(A\text{-}PAH)$ is the total energy of the (relaxed) system consisting of the contaminant atom and the (deformed) PAH molecule, $E_{tot}(PAH)$ is the energy of the isolated PAH molecule in its equilibrium geometry and $E_{tot}(A)$ is the energy of the isolated contaminant atom. ΔE can be divided into the interaction energy between the fragments at their positions and geometries and the energy required for the deformation of the PAH molecule due to this interaction. For small distances between the contaminant atom and the hydrocarbon molecule the interaction energy is positive, giving rise to an energy barrier with a certain shape characteristic for the model.

**3. Results**

**3.1 Convergence of energy and deformation with increasing model size**

In order to justify the use of the results of the cluster calculations to model the interactions between atoms and extended graphene, the energy profiles and especially their maxima must converge for increasing size of the PAH molecule towards graphene, which can be viewed as an infinitely sized PAH molecule. In the following we present simple considerations and tests that give rise to the assumption that this is indeed the case.

a. For larger distances the contaminant atom and the PAH weakly attract each other by dispersion-like interactions that typically scale as or similar to $1/r^6$. For small distances the interaction will be strongly repulsive e.g. as the $1/r^{12}$ in a Lennard-Jones model. In between chemical bonding can occur which is also short-range and directional. All these interactions are short-ranged.
b. Thus, the interaction energy is mainly given by the nearest C atoms, i.e. the atoms of the C-ring in case of permeation at the hollow site at its middle. Since on the other hand the number of C atoms increases with the square of the distance from these C atoms, the interaction energy will approach some constant value if the PAH gets larger. The underlying condition is that the PAH properties relevant for the interaction with incoming particles itself do not change as a property of their size. In case of PAHs this is valid to a good degree, in contrast, for example, to metal nanoparticles where surface bonding interactions are strongly related to their size.
c. The deformation energy should in principle approach a constant value for infinite sized PAH molecules (graphene). Each C atom has 1.5 bonds to neighboring C atoms – three bonds shared between two atoms each. If we employ a simple model where each C-C bond is a harmonic spring with spring constant k, the displacement of one C atom creates a counteracting force proportional to an overall spring constant

$$K = k \left( 1/n + 1/n^2 + \ldots + 1/n^N \right)^{-1}$$

where n is the number of bonds to neighboring C atoms and N is the total number of C atoms constituting the PAH molecule. Since K converges to $k \times (n-1)$ for $N \to \infty$ and $n>1$, the total displacement of a C atom on which a force from an approaching particle is exerted remains finite. Therefore, the energy barrier between atom and PAH will converge to the atom-graphene barrier with increasing PAH size. This can be rationalized by considering that in a small PAH with fixed border atoms larger changes of curvature and bond structure will be induced than in a large one but that this is just compensated by a larger area and number of bonds affected in the large PAH.

In figure 3a the maximum displacement from the original molecular plane is plotted against the number of C atoms of the PAH molecule. The squares give the results for H to Be permeating PAH molecules at the hollow site. The solid lines are functions of the form $f(x)=a(1-e^{-bx})+c$, fitted to the density functional data, where x denotes the number of C atoms and a-c are fitting parameters. One observes a slow convergence of the maximum displacement. This is due to the fact that the boundary region has a large influence in case of small PAH molecules. If instead the bending parameter is plotted as a function of the number of atoms not connected to the boundary (figure 3b), the convergence is faster and the fit functions are better approximations to the data points.

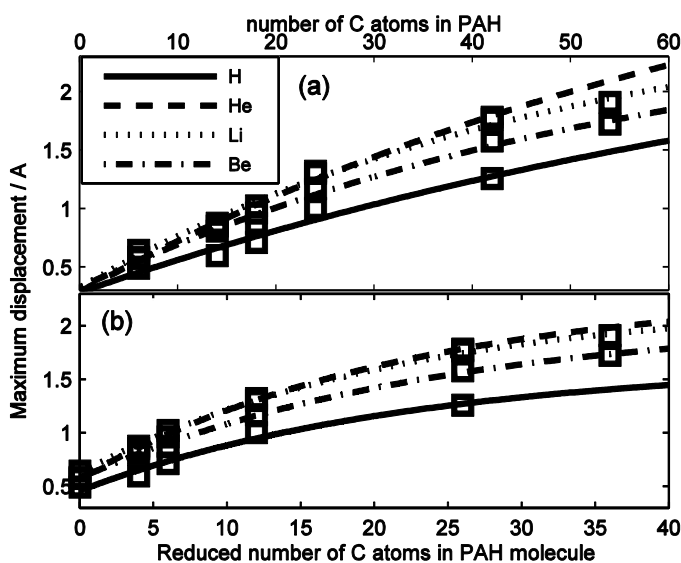

**Figure 3:** Maximum displacement for different atom-molecule interactions plotted versus the number of C atoms in the PAH molecule in (a) and versus a reduced number of C atoms in (b). The better fit in (b) is explained in the text.

From the general considerations given above it follows, that the deformation energy is a quadratic function of the overall deformation of the PAH molecule and the interaction energy reaches saturation very fast with increasing size of the molecules since it depends mainly on the nearest C atoms. Thus, it can be safely assumed that the energy barriers converge much faster than the maximum displacement as a measure for the deformation of the molecules. In fact, it turns out that there exists only slight variations of the barrier height with respect to different sizes of PAH molecules, see figure 4. At least for the three largest PAH molecules under consideration, i.e. coronene, circumpyrene and circumcoronene, the differences in barrier height are within 1 eV ($\approx 1.602 \times 10^{-19}$ J). To investigate how accurate our model is quantitatively compared to DFT calculations using plane waves as basis and introducing periodic boundary conditions to model extended graphene, we compared our results with a DFT study on the interaction of H with graphite surfaces [25]. In the latter work the PW91 functional [45] was used. For a proper comparison we recalculated the energy barrier arising from permeation of H through coronene using this GGA functional. We find that the barrier heights obtained with that functional are 0.5 – 0.8 eV lower depending on the permeation site than those obtained with the PBE0 and B3LYP functionals. The barrier height found using the plane-wave approach and the PW91 functional [25] has been 4.2 eV, whereas the barrier heights found using our cluster approach and the PBE0 and B3LYP functionals have been in the range of 5.2 and 5.7 eV. Taking into account the differences due to the use of different functionals we conclude that a large fraction of this difference can be attributed to the differences between the functionals. It remains, however, difficult to judge from these limited observations if the cluster results can be regarded truly as converged with respect to PAH size. Comparison of additional features of our energy barriers with existing data, i.e. adsorption energies of H and Li, see sections 3.2 and 3.4, and sounding qualitative features in agreement with existing MD potentials in case of H, see section 4, are in favor of our model. We think, that even if our energy barriers are not completely converged, they are already close enough to reality to capture the essential physics needed for proper dynamical studies and can be used to inform potential building or re-check already existing potentials.

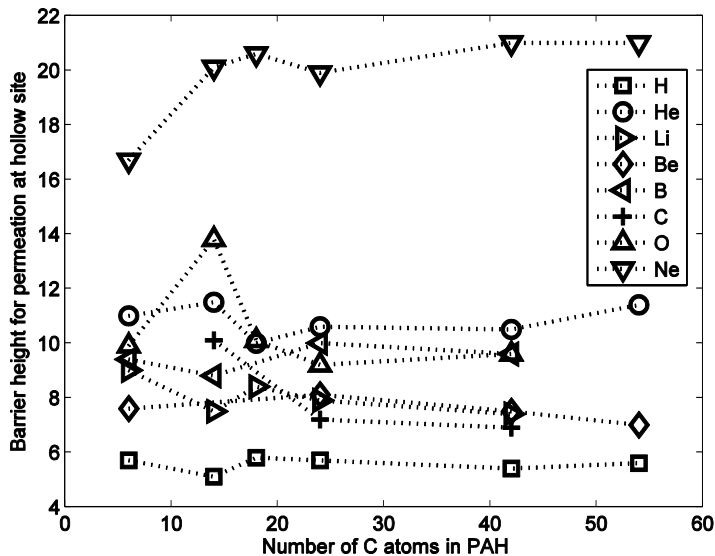

**Figure 4:** Barrier height versus number of C atoms contained in the different PAH molecules for permeation at hollow site for the different atomic species under consideration.

**3.2 Adiabatic permeation at hollow site**

Our calculations correspond to the physical situation in which the surface structure relaxes much faster than the contaminant atom approaches it. One can call this the adiabatic interaction regime. Two other extreme cases, the in-plane regime as well as the ultra-fast regime, are also considered and will be discussed in detail in section 3.5. In the following we discuss the energy barriers and deformation effects for adiabatic permeation of the atoms H up to Ne through different PAH molecules as models for graphene at hollow site. The energy barriers for adiabatic permeation through the three largest PAH molecules (coronene, circumpyrene and circumcoronene) at a hollow site are shown in figure 5. The energy barriers can roughly be divided according to their shape into 'regular' ones and more complex (or 'irregular') ones. The regular energy barriers are characterized by a steep increase of the energy at about 2 Å distance between the contaminant atom and the PAH molecule under consideration up to a maximum in energy and a steep decrease (jump) afterwards back to about zero energy. H, He, Li (concerning the results obtained with B3LYP), Be and Ne were interacting with the PAH in this 'regular' way. The 'irregular' energy barriers are discussed individually for the various types of contaminant atoms. The characteristic energy maxima for permeation through coronene (if available) are summarized in table 1. The maxima, i.e. the heights of the energy barriers, approximately follow the order of the atoms in the periodic table. Exceptions are He with a maximum of about 12 eV, which is due to its closed shell configuration, and C and O. Their ability to form chemical bonds with the surface lowers the barrier height for most distances, compared to inert atoms of similar van-der-Waals radii. Except for these atoms, the energy maxima range from about 5 eV in case of H up to about 20 eV for Ne. The energy barrier for H is in reasonable agreement with an earlier DFT study [25], as already discussed in section 3.1. Studies on the adsorption of Li at graphite surfaces [46-48] allow for a comparison of our results in front of the permeation barrier. Binding energies of adsorbed Li atoms at a hollow site of earlier works are in a range of 0.17 – 1.70 eV and adatom distances are in a range of 1.63 – 2.1 Å depending on the underlying methodology and used model chemistry [47]. The most recent reported values of 1.21 eV and 1.84 Å were obtained using Car-Parinello molecular dynamics based on the density functional theory using the PBE functional and periodic boundary conditions to simulate the infinitely extended graphite (0001) surface [46]. Our cluster model yields adsorption energies in a range of 0.27 – 1.32 eV and adatom distances of 1.6 – 1.8 Å, where the results obtained with the PBE0 functional are in a range of 0.78 – 1.32 eV and thus much closer (than the B3LYP values) to the results of the works intended to accurately predict this binding energy. From a comparison with earlier theoretical work based on the B3LYP functional as well but using larger basis sets [48], we conclude that the 6-31G basis set is too small to yield accurate predictions of adsorption energies in this case. However, we note that in the region of the permeation barrier, on which this paper focuses, the two functionals yield very similar results, indicating that the lower quality of the results obtained with B3LYP in front of the barrier vanishes when the contaminant atom permeates the PAH molecule.

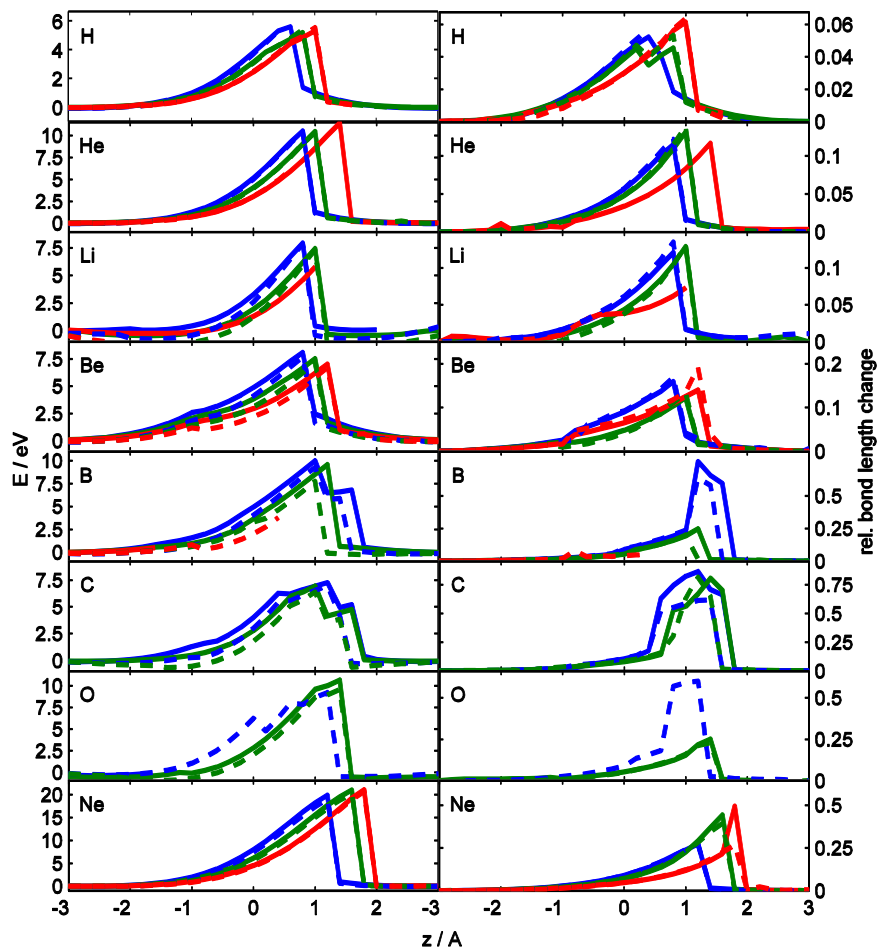

**Figure 5:** Energy barriers (left) and relative change in C-C bond length in the vicinity of the permeation site (right) for the permeation of H to C, O and Ne through coronene (blue), circumpyrene (green) and circumcoronene (red). Results are shown for the B3LYP (solid lines) and PBE0 (dashed lines) density functionals.

In addition to the energetics we also tried to quantify the deformation of the PAH molecule. As a simple measure, the relative change in the C-C bond length in the vicinity of the permeation site is calculated as a function of the z-coordinate of the contaminant atom (figure 5.). The 'irregularities' in the energy barriers are mostly reflected in the deformation curves. Values for the maxima are summarized in table 1. Also the maximal deformation follows approximately the order of the atoms in the periodic table (except for C and O, see the discussion below). The values range from small changes of 5-6% for H up to about 30% in case of Ne and even up to 80-90% in some cases that are discussed below.

**Table 1:** Energy maxima, $E_{max}$, and maxima of the relative change of the C-C bond length, $D_{max}$, for the permeation of H to Ne through coronene. The values correspond to the two functionals B3LYP/PBE0.

| Atom | H | He | Li | Be | B | C | O | Ne |
|---|---|---|---|---|---|---|---|---|
| $E_{max}$/[eV] | 5.7/- | 10.6/10.4 | 7.9/7.8 | 8.1/7.6 | 10/9.3 | 7.2/6.9 | -/9.2 | 19.9/19 |
| $D_{max}$/[%] | 5.2/- | 11.5/12.2 | 12.1/13.5 | 16/16.8 | 75[a]/64[a] | 86/60 | -/60 | 27/27 |

[a] large values due to asymmetric deformation, see text; characteristic values for symmetric deformations are about 20%

In case of B the results from the two functionals differ most, about 20% at the energy maxima in case of permeation through circumpyrene. A pronounced maximum in the deformation curve for permeation through coronene can be seen which stems from an asymmetry in the way this PAH is deformed, see figue 6. This asymmetry is also responsible for the unusual 'buckle' in the respective energy curve (figure 5).

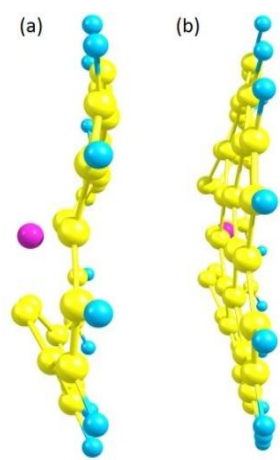

**Figure 6:** Difference between permeation of B (violet sphere) through (a) coronene and (b) circumpyrene. The asymmetrical deformation caused by bond formation can be seen in (a).

For C and O, energy barrier as well as deformation curves exhibit more complexity. Both energy and deformation curves reveal relative kinks, buckles and additional minima occurring in front of the steep increase to the energy maximum. This is a consequence of the strong chemical reactivity of these elements with the C atoms and show the formation and breaking of chemical bonds. This chemical activity in the permeation region can partly be analyzed by decomposing the total energy barrier into the energy that is stored in the deformation of the PAH molecule and the remaining interaction energy, which can be done by calculating the energies of the deformed PAH molecules at each scan step. We carried out this decomposition for coronene, see figure 7. One observes that the systems that give rise to 'irregular' curves are those which are attractive in the permeation region of the potential. This happens in the B-, C-, and O-PAH systems. For all the other systems under consideration the interaction energy contribution is always positive.

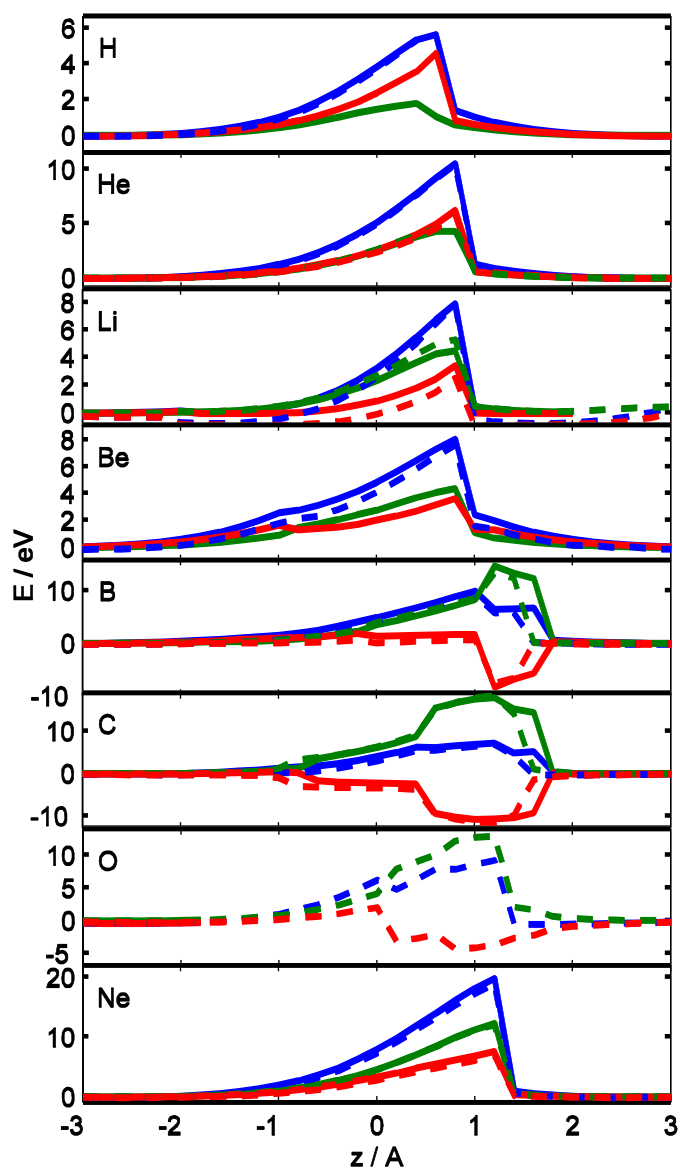

**Figure 7:** Decomposition of total energy barrier (blue) into 'pure' interaction energy (red) and deformation energy (green) for the atoms H-C, O and Ne permeating coronene. Results are shown for B3LYP (solid lines) and PBE0 (dashed lines).

**3.3 Adiabatic permeation at a bridge site**

For H, He, Li and Be the interaction at bridge and top sites has also been investigated. The total energy barriers and relative changes in C-C bond length in the vicinity of the permeation site are quite similar to the ones discussed above. The differences that arise from the higher energy requirement to break a C-C bond compared to the one to enlarge a hexagonal opening are discussed here.

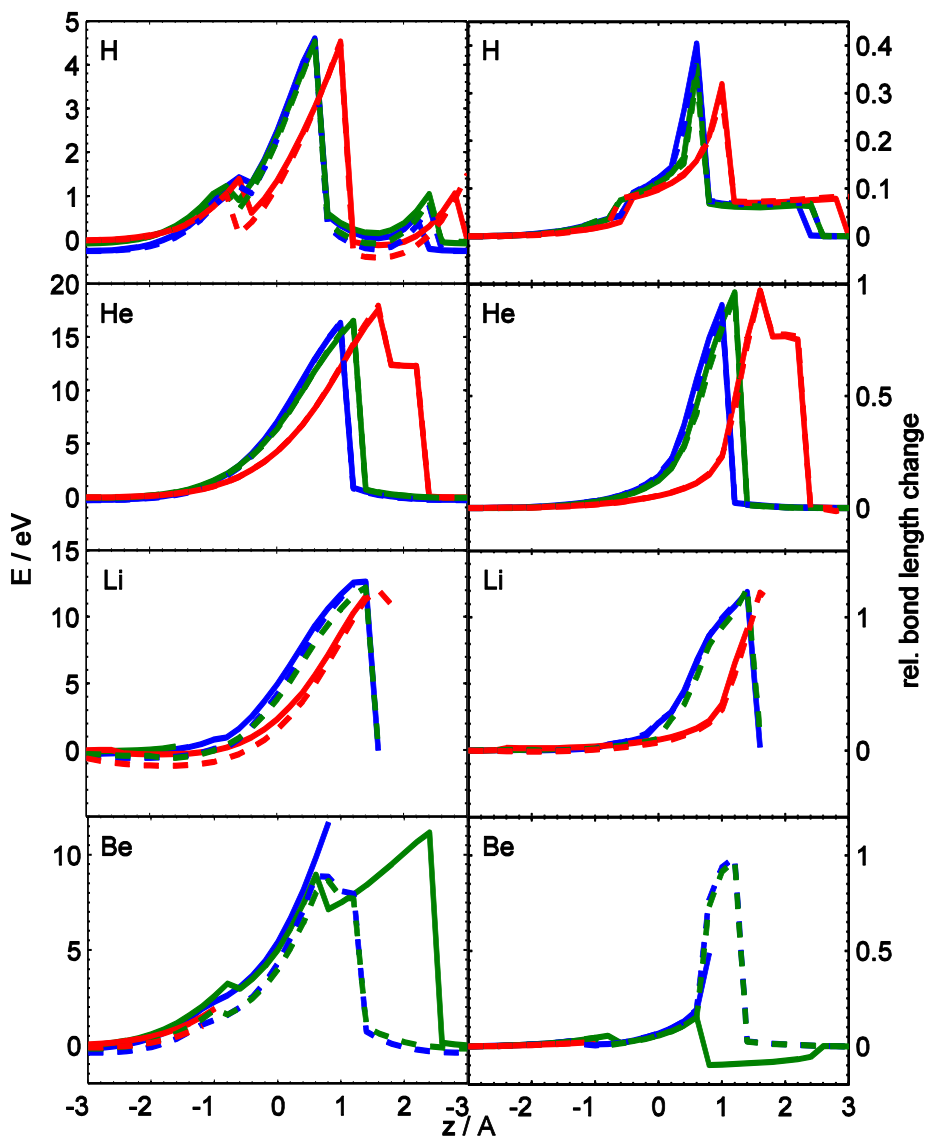

**Figure 8:** Energy barriers and relative change in C-C bond length in the vicinity of the permeation site for adiabatic permeation through anthracene (blue), coronene (green) and circumpyrene (red) at a bridge site for H, He, Li and Be. Results are shown for B3LYP (solid lines) and PBE0 (dashed lines).

In figure 8 the results of the calculations for H, He, Li, Be and the three PAH molecules are given. Concerning H, B3LYP and PBE0 yield similar results as in case of permeation at the hollow site. The important difference, however, are the two additional maxima in both curves. The closer approach of H to the C atoms causes a tetrahedral arrangement when H approaches to about 1 Å, see figure 9. The same happens when H leaves at the other side of the PAH where the tetrahedral configuration is maintained until z=2.2-2.8 Å, leading to the energy maximum in that region. Simultaneously with the C-H bond formation the C-C bond is weakened and the C-C distance becomes much larger. The energy maximum is almost equal for all three kinds of PAHs. The heights of these maxima (about 4.5 eV) are smaller than the corresponding maxima for the hollow site case (about 5.7 eV). One observes that the penetration of H through the PAH is mediated by C-H bond formation. This is only the case for H out of the four low-Z atoms H to Be.

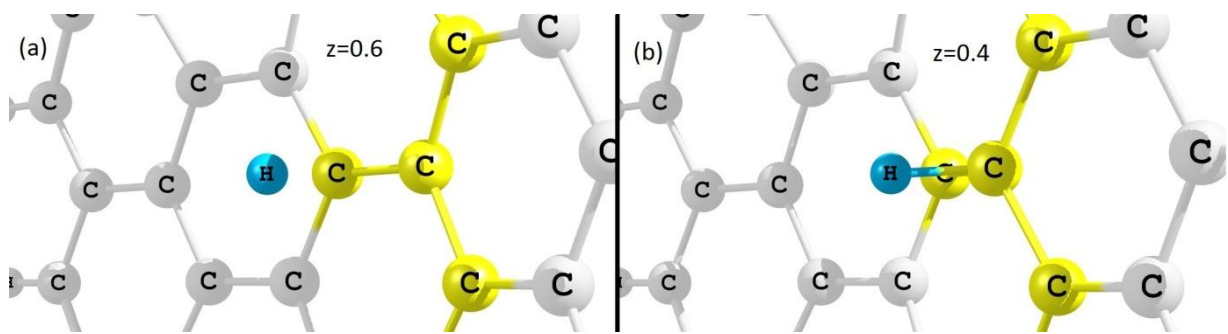

**Figure 9:** Tetrahedral configuration in the vicinity of the permeation site in the middle of the two C atoms when the H atom approaches z=0.4 Å shown in (b) in contrast to the planar configuration in (a) at the distance at z=0.6 Å of the H atom to the original molecular plane.

It can be observed easily that the agreement between the two methods/basis sets is very good in the case of He permeating at bridge site. Since for He repulsion dominates by far, large values of both the total energy (about 16-18 eV) as well as of bond elongation (up to 100%) can be observed.

We note that the shape of the curves seems to depend somewhat on the choice of PAH molecule. Whereas both curves are very similar to those obtained in case of hollow site for anthracene and coronene, the curves for interaction between He and circumpyrene have a different shape. In the latter case, the PAH molecule does not flip back into its original plane after permeation of He, but the two central C atoms remain close to He and are indeed following the leaving He atom a bit until the point, where the PAH molecule flips back after all into its original geometry. A reason for this difference may be the greater deformability of circumpyrene due to smaller boundary effects.

The potential energy curves for the interaction of Li with the three hydrocarbons are very similar to those obtained for the case of He. The maximum values of the total energy are about 12 eV for the two functionals. The relative change in C-C bond length of the central C-C bond is again rather large, up to 120%.

The most interesting features concerning Be permeating through PAH molecules at the bridge site are the differences between the results obtained with the B3LYP and PBE0 functionals. It can be observed that in case of the coronene molecule both the total energy curve as well as the curve concerning change in C-C bond length differ completely. The calculations with different methods also result in a different behavior of coronene when Be permeates it. In case of PBE0 the Be atom just moves between the two C atoms forming the central C-C bond, whereas in case of B3LYP the Be atom pushes aside this central bond together with both C atoms as a whole, see figure 9. In the former case, this results in an energy maximum at about 8.5 eV and large separation of the two considered C atoms, reflected by the relative change of bond length of about 100%. In the latter case, an even larger energy maximum at about 11 eV is found at z=2.4 Å, together with a very small extension or even shortening of the central C-C bond. In both cases, coronene reestablishes its original geometry at the end. Missing convergence in the SCF procedure at z=0.8 Å prevented us to compare with circumpyrene and anthracene for distances larger than 0.8 Å after permeation. The rise of C-C bond length change for that case, at least, indicates that here the behavior is comparable to the results of PBE0, however, accompanied by a larger energy maximum greater than or equal to about 12 eV.

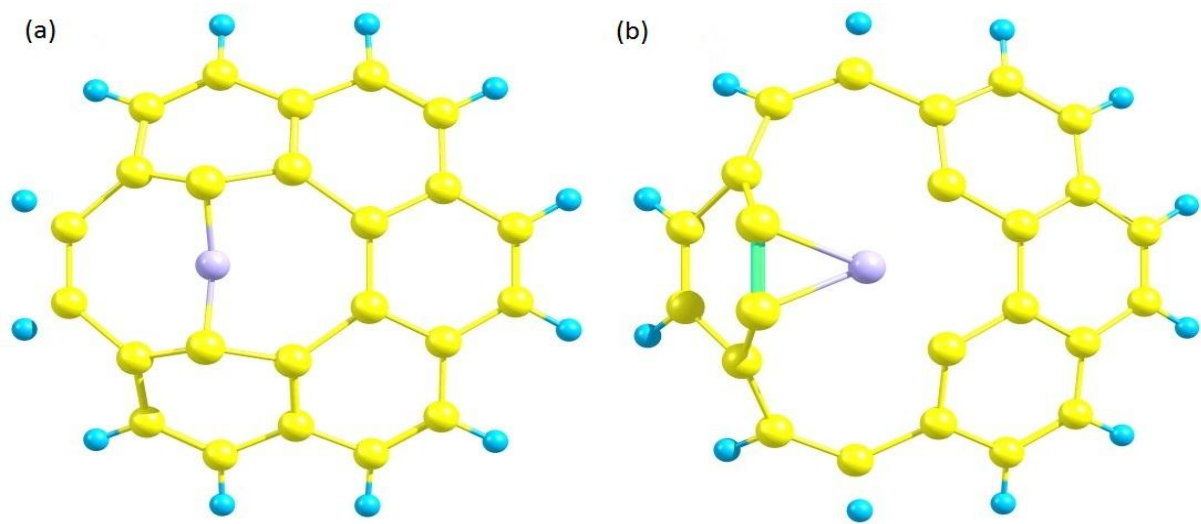

**Figure 9:** Deformation of coronene due to permeation of Be at bridge site, once obtained with (a) PBE0 and once with (b) B3LYP.

**3.4 Adiabatic permeation at top site**

In the third geometry investigated the contaminant atom is located on top of a C atom. The change of the length of one C-C bond between the C at the interaction site and one of its neighbors with the largest deviation from undisturbed bond length has been chosen as a measure of the PAH distortion. In general, permeation at top site leads to more repulsive energy curves as the other geometries and can be imagined to proceed as follows: The contaminant atom first pushes the C atom at the permeation site simply out of the undisturbed molecular plane. Due to physical or purely numerical asymmetries the C atom starts to be pushed aside as the z-coordinate of the contaminant atom is increased. As the contaminant atom leaves the deformed PAH molecule, the C atom flips back into the original position.

We start with the results for the H-PAH system (figure 10). As for the permeation at bridge site there exist additional minima in energy as well as non-monotonous bond length changes. These minima are less pronounced than in the case of permeation at bridge site but like there they are caused by C-H bond formation. The results from the functionals do not agree as well as for the other geometries. The respective maxima are 8 eV for PBE0 and 9 eV for B3LYP, both at about 0.6 Å. The relative change in C-C bond length is 60-80%. The small minima in front of the barrier (at about z=-1.4 to -1.2 Å) correspond within 0.2 eV to activation barriers and adsorption minima discovered by earlier DFT studies [24-26, 49] and experiment [50] which focused on chemisorption. The C atom extrudes of the PAH plane and a tetrahedral configuration is formed with its neighbors and the approaching H atom.

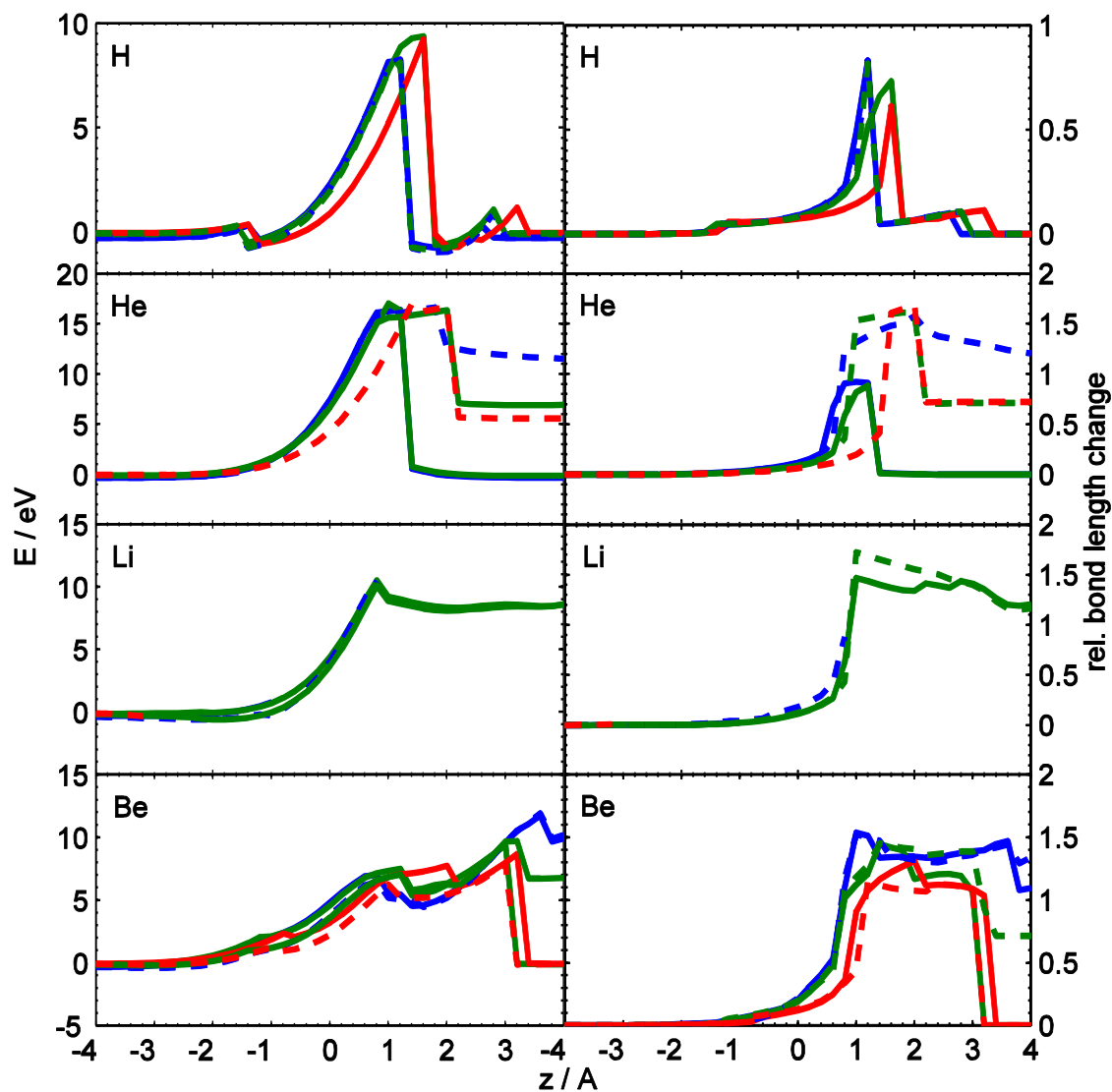

**Figure 10:** Energy barriers and relative change in C-C bond length in the vicinity of the permeation site for adiabatic permeation through anthracene (blue), coronene (green) and circumpyrene (red) at the top site for H, He, Li and Be. Solid lines are for B3LYP and dashed lines for PBE0.

Concerning the permeation of He through PAH molecules at top site, for PBE0 the PAH molecules do not relax back into their original geometrical structure. In terms of the energy barrier this leads to a larger total energy at the end of the scan (figure 10). The reason for this is that the central C atom in anthracene remains apart from the original plane while coronene and circumpyrene change their structure (figure 11). Three coronene hexagons convert into two pentagons and one heptagon while in circumpyrene the four central hexagons are replaced by two pentagons and two heptagons. This isomerization gives also rise to a change in the electronic structure and material properties [34] and is prototypical for a beginning degradation of the surface material.

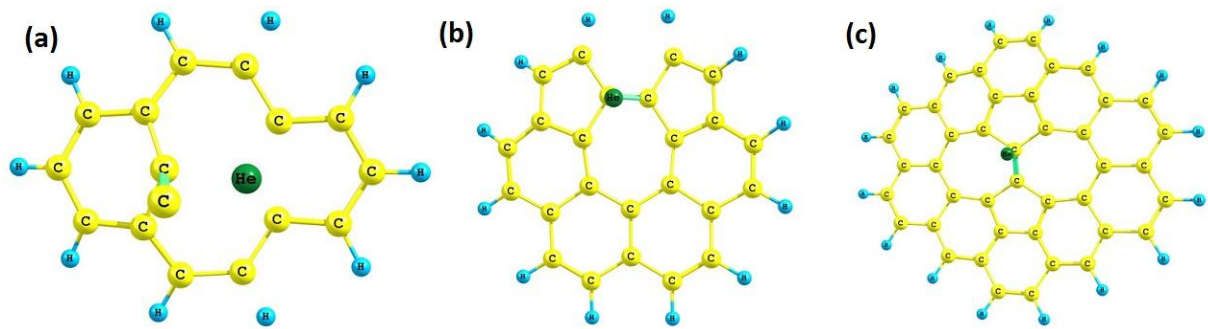

**Figure 11:** Remaining geometrical structures of anthracene (a), coronene (b) and circumpyrene (c) after permeation of He (dark green) through the molecule at top site.

In case of Li as contaminant atom the PAH molecules do not flip back after permeation of Li at top site but the C atom that has been pushed aside by the incoming Li atom instead follows the Li atom, which further increases the deformation. Consequently, the C-C bond length changes by up to 170% (figure 10).

For Be, the situation is even more complicated (figure 10). The reason for the two pronounced maxima in energy and rectangular shapes can be rationalized as follows: The central C atom is pushed out and to the side. When this C atom flips partially back, the energy of the system is reduced, but this C forms a C-C bond perpendicular to the original PAH plane. As the contaminant atom leaves the PAH, this out-of-plane bond can return into the original molecular plane. This process is representative for the energy profiles shown in figure 10 even though they differ in details from each other.

We conclude this section by summarizing the values for energy maxima and the maxima in relative change in C-C bond length in Table 2. We note that permeation of atoms through coronene requires least energy at hollow site, except for H, which permeates most easily at bridge site. The energy maxima differ substantially for the case of He. There the bridge and top site maxima are about 6-7 eV larger in value than at hollow site. For Li and Be the values are rather similar and the differences between the different sites are about 2-3 eV. The value for H at top site is comparable to those of Li and Be, and substantially (4-5 eV) larger than the values at bridge and hollow site. The maxima in relative change in C-C bond length are very small at hollow site (5-20%) compared to those at bridge and top site, where they are around 100%. Isomerizations of the PAHs are found, for example, in case of He at top site for PBE0 and in case of Be at bridge site for B3LYP.

**Table 2:** Values for energy maxima, $E_{max}$ [eV] and maxima for relative change of C-C bond length, $D_{max}$ [%] for permeation through coronene of the atoms H to Be at the three high symmetry sites hollow, bridge and top. Values are given for both methods in use in the following order: B3LYP/PBE0.

|  | H | He | Li | Be |
|---|---|---|---|---|
| $E_{max}$ at H site | 5.7/- | 10.6/10.7 | 7.9/7.8 | 8.1/7.6 |
| $E_{max}$ at B site | 4.5/4.5 | 16.6/16.8 | -/12.3 | 11.2/8.6 |
| $E_{max}$ at T site | 9.4/8.1 | 17.1/16.4 | 10.6/10.2 | 9.4/9.8 |
| $D_{max}$ at H site | 5.2/- | 11.5/12.2 | 12.1/13.5 | 16/16.8 |
| $D_{max}$ at B site | 73/82 | 94/97 | -/122 | 15/97 |
| $D_{max}$ at T site | 73/82 | 89/161 | 147/172 | 144/145 |

### 3.5 Comparison of different regimes

As outlined in section 3.2, in addition to the adiabatic interaction regime, two other regimes were investigated. In the 'in-plane regime' the PAH is restricted to stay planar. This relates to a model of graphene that is restricted to planarity by external forces. In the 'ultra-fast regime' no relaxation of the PAH is allowed at all, modeling thereby their behavior upon impact of atoms with very high velocities. Since both restrictions lead to much closer approaches between the contaminant atom and C atoms of the PAH molecules, only permeations at the hollow site make sense and have been investigated. We restricted our calculations to coronene as well. By definition, the energy barriers and deformation curves for these two regimes are symmetric with respect to z=0, except for small numerical artifacts and no 'hysteresis' occurs. Energies and C-C bond length changes in all three regimes are shown in figure 12.

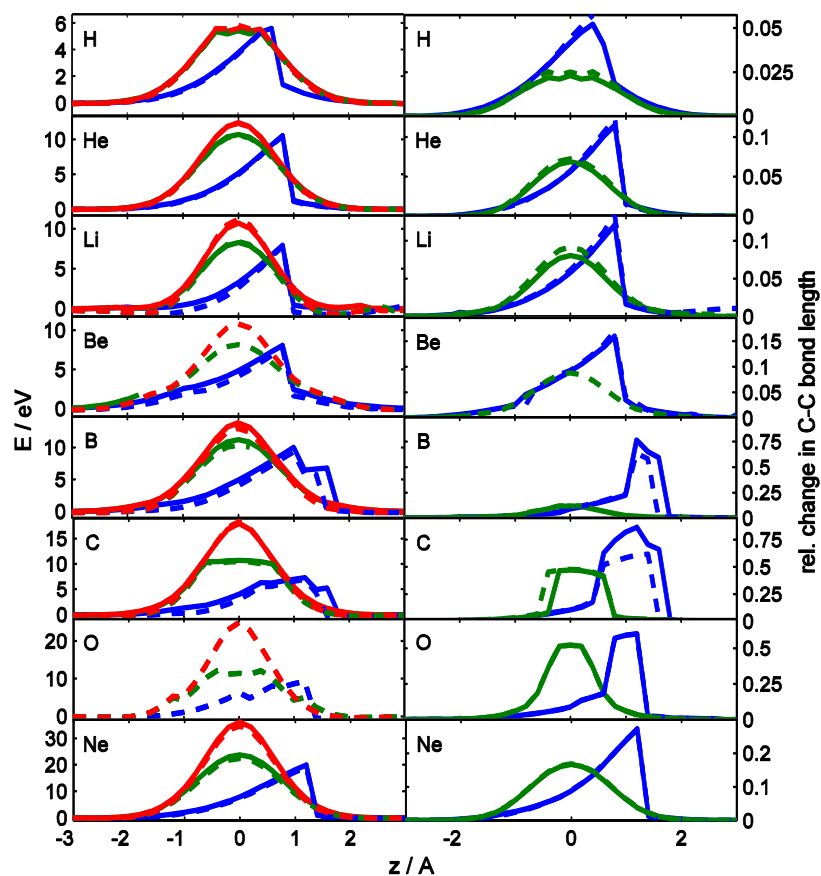

**Figure 12:** Energy barriers (left) and deformation curves (right) for the adiabatic (blue), in-plane (green) and ultra-fast (red) interaction regimes. Results are shown for B3LYP (solid lines) and PBE0 (dashed lines).

Interestingly, the heights of the planar and adiabatic energy profiles are quite similar, except for C and O. This indicates that the out-of-plane stabilization energy is small. Obviously, the contrary is true for the relative change in C-C bond length in the vicinity of hollow site. There, the planar part is lower, about 1/2-2/3, of the adiabatic change in bond length. Both functionals yield very similar results. The energy and deformation curves for the in-plane and ultra-fast regimes are characterized by their maxima (table 3). The dependence of the barrier height on the atomic number for the three regimes discussed in this paper is depicted in figure 13. Though not discussed in this paper, as they are expected to be less relevant for nuclear fusion, the values for N and F are given in table 3 and figure 13 as well for completeness.

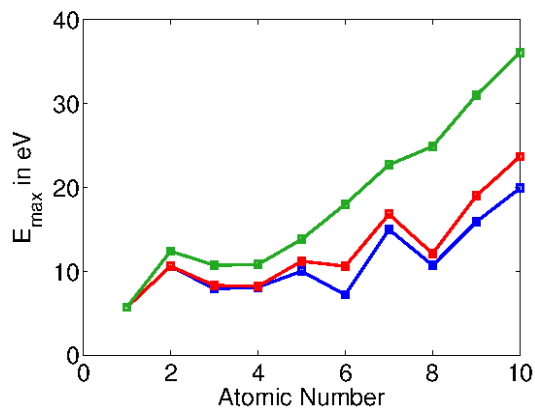

**Figure 13:** Dependence of the energy barrier height on the atomic number of the atom permeating through coronene at hollow site for the adiabatic (blue), in-plane (red) and ultra-fast (green) regime.

**Table 3:** Values of the energy maxima, $E_{max}$ [eV] and maxima for relative change of C-C bond length, $D_{max}$ [%] for permeation of H to Be through coronene at hollow site for in-plane (top) and ultra-fast (bottom) interaction regime. Values are given for the B3LYP/PBE0 functionals.

|     | H | He | Li | Be | B | C | N | O | F | Ne |
|---|---|---|---|---|---|---|---|---|---|---|
| $E_{max}$ | 5.7/5.6 | 10.6/10.7 | 8.3/8.4 | -/8.2 | 11.2/10.4 | 10.6/10.4 | 16.8/- | -/11.3 | -/19 | 23.7/22.4 |
| $D_{max}$ | 2.3/2.5 | 6.8/7.2 | 8.0/9.1 | -/8.8 | 11.9/12.3 | 47/48 | 12.7/- | -/52 | -/35 | 17/17 |
| $E_{max}$ | 5.7/5.9 | 12.4/12.4 | 10.7/11.2 | -/10.8 | 13.8/13 | 18/18.4 | 22.7/22 | -/25 | 30.8/29.5 | 36.1/34.5 |

### 3.6 Hydrogenated surfaces

We also explore the adiabatic interaction of these atoms with the 100% hydrogenated version of coronene as a model for graphane (CHC) [38]. The hollow interaction site is still well defined as the midpoint of one of the C hexagons which are now puckered. This system is important for fusion research since strongly hydrogenated graphite surfaces occur in fusion experiments using graphite plasma facing component (PFC) materials [5]. Indeed, it can be expected that H-covered surfaces are often more abundant than bare ones.

For C and O no suitable results could be obtained due to missing convergence in the SCF procedure during the geometry optimizations. The atoms most relevant concerning nuclear fusion H, He and Be as well as B, Li and Ne have been investigated, however.

As in the case of adiabatic permeation through PAH molecules we can decompose the results for energy barriers and deformation curves in simple ones for H, He, Li, and Ne and more complex ones for Be and B. In the former case the total energy barriers and deformation curves are very much like those for permeation through PAH molecules concerning their shape, with slightly different values of energy barriers and deformation maxima. It turns out that the barriers are slightly reduced, compared to the pure PAHs while the C-C bond length changes are about the same The results are depicted in figure 14.

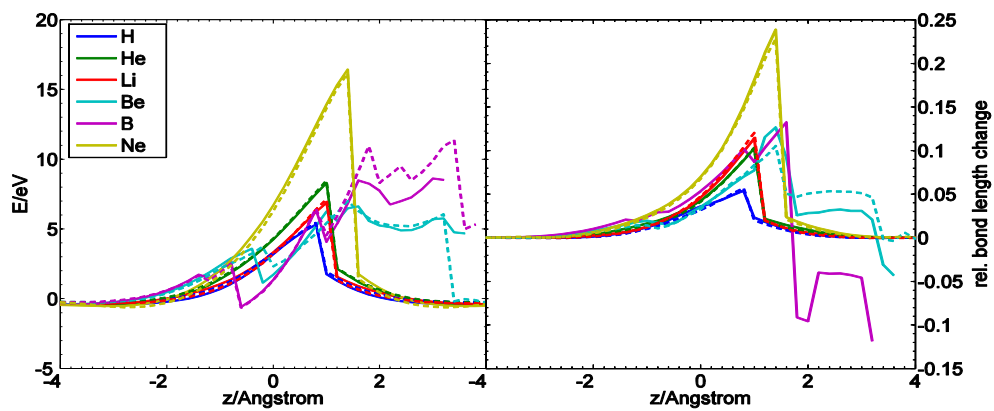

**Figure 14:** Energy barriers (left) and deformation curves (right) for adiabatic permeation of H (blue), He (green), Li (red), Be (turquoise), B (violet) and Ne (gold) through 100% hydrogenated coronene. Results are shown for B3LYP (solid lines) and PBE0 (dashed lines).

The situation is different when Be and B are contaminant atoms since they form bonds to the CHCs. Their energy barriers and deformation curves are complex and zagged but are similar to each other. In both cases, we observe three maxima in the energy barriers correspond to the three layers (H, then C, then H again) the atoms are permeating through. The additional small maxima visible in some of these curves are thought to be numerical artifacts. In both cases of Be and B permeating through CHC we observe that a $H_2$ molecule is released from the C-H-cluster when the contaminant atom permeates hydrogenated coronene. This is an indication that the bombardment of hydrogenated surfaces can lead to energetically driven secondary molecule formation.

## 4. Conclusions

We investigated the interaction of low-Z atoms H up to C, O and Ne with PAH molecules as models for graphite (0001) surfaces by means of DFT methods using the B3LYP and PBE0 functionals. From these calculations we derived energy barriers and curves characterizing the deformation of the PAH molecules for permeation of the atoms under consideration through the hydrocarbon molecules at different high-symmetry sites. Overall, both functionals gave similar results. Comparison with earlier theoretical and experimental work on the interaction of H with graphite and adsorption of Li at graphite (0001) surfaces results in a reasonable agreement. We conclude that an accuracy of about 10-15% concerning the barrier height can be achieved with these functionals in conjunction with the small split-valence basis sets 3-21G and 6-31G. By investigating the convergence of the energy barriers and comparison to earlier work based on DFT using plane waves as basis and periodic boundary conditions, we conclude that the geometric and energetic characterization of atomic interaction with graphite surfaces can be done on a quantum chemical level by investigating the atomic interaction with PAH molecules of at least the size of coronene. Especially for the close relation to nuclear fusion concerns, we also investigated atomic interaction with 100% hydrogenated coronene. We found that in this case the energy barriers heights are slightly lower than for pure coronene. Furthermore, we found a purely energetically driven mechanism leading to the formation of molecules not incorporating the impact atom (secondary molecule formation).

Though our results are surely limited in accuracy (by the 15% bound given above), they are thought to represent the essential physics underlying projectile-surface interactions as those relevant in the chemical erosion of graphite used as divertor material in the next-step fusion reactor ITER. In fact, the barrier heights derived for H impinging on graphite (0001) surfaces at the three high-symmetry sites hollow, bridge and top account for the importance of swift chemical sputtering [51] and the mechanisms underlying the graphite peeling [21]. The reason underlying these two consequences is that the barrier for H is lowest not for permeation at the hollow site but at bridge site. An H atom approaching graphite is thus directed towards the lowest barrier maximum, i.e. to the boundaries of the hexagonal cycles constituting the graphite lattice. That this is exactly the case can be made obvious by performing quantum dynamical studies using potentials based on potentials constructed using our methodology [52]. This has certain consequences concerning the other possible projectile atoms in nuclear fusion as well. He, Li and Be all face the smallest barrier maximum at the hollow site. Molecular dynamics simulations have revealed that small concentrations of noble gases in the projectile species only negligibly influence the sputtering yield [27]. This can be interpreted easily with the help of the He-graphite potential. In contrast to H, He is directed towards the centers of the aromatic cycles when impinging on the graphite surface. He does thereby not contribute to the erosion by swift chemical sputtering and in addition partitions its energy over the six C atoms constituting the respective hexagon rather than the edge of the hexagon as in case of H, see again [52]. Based on these reasoning we conclude that the interaction of Li and Be can be characterized in a similar manner and contribute to the sputtering of graphite (0001) surfaces mainly via physical sputtering for higher impact energies than via chemical sputtering as does H. For the chemical erosion of ideal graphite (0001) surfaces due to H bombardment we conclude further, that the main pathways for the degradation of the surface material are surface-chemisorption at lower energies (up to 1 eV) and swift chemical sputtering at slightly elevated energies (up to 4 – 5 eV). The first pathway corresponds to the adsorption minima in the H-graphite potential at top site, whereas the second one corresponds to the barrier height. An H atom approaching the surface with about 5 eV will be completely stopped when climbing up to potential barrier and thus maximizes the time of interaction with the graphite lattice. As the probability for swift chemical sputtering depends on the time the H atom stays in the interaction region [51], an impact energy of about 5 eV corresponds to a high probability for adsorption of H projectiles. Molecular dynamics simulations have shown that this is indeed the case [18]. We note that this implies that there is nothing new in our H-graphite potentials, but that they are in good agreement with earlier observations and capture the essential physics. Based on that we are sure that our energy barriers for the other atomic species can serve as a good basis to re-check (or inform new) molecular dynamics potentials to study the sputtering of graphite surfaces in a mixed material scenario as it will be encountered in the next-step fusion reactor ITER. He and Ne can be regarded as prototypes for noble gases and this is well reflected by the similarity of their energy barriers for permeation through our models for graphite (0001) surfaces. The energy maxima of these barriers are thus believed to be good parameters for the construction of two-body potentials describing the essential physics underlying their influence on the sputtering of graphite (0001) surfaces. Such two-body potentials can then be relatively easy adjusted to model the interaction of Kr and Xe with graphite as well. The dynamics of Be bombardment of graphite should as outlined above resemble that of He, since the characteristics of the energy barriers are similar (lowest at hollow site, largest at top site, in between at bridge site). Thus the influence of Be on the sputtering yield should be quite small as well in a mixed material scenario. Atomic C plays a role due to self-sputtering and well-parameterized molecular dynamics potentials are needed to model this on a physically good basis. Although being in principle in a vacuum chamber, a certain amount of O can never be avoided and the influence of O on the sputtering dynamics might become interesting. On balance, one has to face a multi-species scenario when modeling the sputtering dynamics of graphite surfaces. These species will of course not only consist of atomic and ionic species but molecular species as well. A DFT study of small hydrides is thus foreseen for the near future [53].

An additional feature of our investigation has been the discussion of the three different interaction regimes referred to as adiabatic, in-plane and ultra-fast interaction regimes. Especially interesting is the result that the adiabatic and in-plane regime yield barrier heights that are very close to each other (although the shape of the curves are obviously different due to symmetry). Physically that means that the energy of the system cannot be lowered significantly by a planar relaxation of the surface. In graphite, the distortion of the surface layer into the direction of the next layer is somewhat hindered by an interlayer potential, such that the total physical potential for permeation of the surface lies in between the adiabatic and in-plane regime. However, this has no great influence on the barrier height as our results indicate, which is thought to be an useful information for molecular dynamics potential builders. Finally, the ultra-fast regime barriers show that the strength of the repulsion depends on the dynamics, i.e. on the velocity with which the atomic projectiles approach the surface. As the reaction on the atomic scale is thought to be a very fast process, this might play a role only for highly energetic particles that – on the other hand – are already far beyond the barrier maximum. In principle, however, the true projectile-surface potentials is located between these three limiting conditions, but for applications in modeling the chemical erosion of graphite, which is done typically in an energy range of 1 – 10 eV, the relevant interactions can be safely assumed to be far from the ultra-fast interaction regime.

**Acknowledgements**

This work, supported by the European Commission under the Contract of Association between EURATOM and the Austrian Academy of Sciences, was carried out within the framework of the European Fusion Development Agreement. The views and opinions expressed herein do not necessarily reflect those of the European Commission. This work was supported by the Austrian Ministry of Science BMWF as part of the Uni-Infrastrukturprogramm of the Forschungsplattform Scientific Computing at LFU Innsbruck. Support from the DK+ on computational interdisciplinary modeling and from the RFBR-FWF project 09-03-91001-a is also gratefully acknowledged.